\documentclass[journal=jctcce,manuscript=article]{achemso}
\usepackage[version=3]{mhchem} 
\usepackage{placeins}
\usepackage{hyperref}
\usepackage[authormarkuptext=name,authormarkup=none]{changes}
\definecolor{acolour}{RGB}{0, 0, 255}
\definecolor{red}{RGB}{255, 0, 0}
\definechangesauthor[name={Yorick}, color={acolour}]{YS}
\author{Yorick L. A. Schmerwitz}
\affiliation[University of Iceland]
{Science Institute and Faculty of Physical Sciences, University of Iceland VR-III, 107 Reykjavík, Iceland}
\alsoaffiliation[Julius-Maximilians-Universität Würzburg]
{Institut für Physikalische und Theoretische Chemie, Fakultät für Chemie und Pharmazie, Julius-Maximilians-Universität Würzburg, Emil-Fischer-Straße 42, Würzburg, D-97074, Germany}
\author{Vilhjálmur Ásgeirsson}
\affiliation[University of Iceland]
{Science Institute and Faculty of Physical Sciences, University of Iceland VR-III, 107 Reykjavík, Iceland}
\author{Hannes Jónsson}
\affiliation[University of Iceland]
{Science Institute and Faculty of Physical Sciences, University of Iceland VR-III, 107 Reykjavík, Iceland}
\email{hj@hi.is}
\affiliation[University of Iceland]
{Science Institute and Faculty of Physical Sciences, University of Iceland VR-III, 107 Reykjavík, Iceland}
\title{Improved Initialization of Optimal Path Calculations Using Sequential Traversal over the Image Dependent Pair Potential Surface}
\begin{document}
\begin{abstract}
In reaction path optimization, such as the calculation of a minimum energy path (MEP) between given reactant and product configurations of atoms, it is advantageous to start with an initial guess where close proximity of atoms is avoided and bonds are not unnecessarily broken only to be reformed later. When the configurations of the atoms are described with Cartesian coordinates, a linear interpolation between the endpoints can be problematic, and a better option is provided by the so-called image dependent pair potential (IDPP) approach where interpolated pairwise distances are generated to form an objective function that can be used to generate an improved initial path. When started with a linear interpolation, this method can, however, still lead to unnecessary bond breaking in, for example, reactions where a molecular subgroup undergoes significant rotation. In the method presented here, this problem is addressed by constructing the path gradually, introducing images sequentially starting from the vicinity of the endpoints while the distance between images in the central region is larger. The distribution of images is controlled by systematically scaling the tightness of springs acting between the images until the desired number of images is obtained and they are evenly spaced. This procedure generates an initial path on the IDPP surface, a task that requires insignificant computational effort, as no evaluation of the energy of the system is needed. The calculation of the MEP, typically using electronic structure calculations, is then subsequently carried out in a way that makes efficient use of parallel computing with the nudged elastic band method. Several examples of reactions are given where the linear interpolation IDPP (LI-IDPP) method yields problematic paths with unnecessary bond breaking in some of the intermediate images, while the sequential IDPP (S-IDPP) method yields paths that are significantly closer to realistic minimum energy paths.
\end{abstract}
\section{Introduction}\label{introduction}
Theoretical studies of the mechanism and rate of atomic rearrangements such as chemical reactions, diffusion events, and conformational changes, often involve finding an optimal transition path. Typically, this path is taken to be a minimum energy path (MEP) between given configurations of the atoms corresponding to initial and final state minima on the energy surface. The MEP illustrates the transition mechanism, and the highest rise in the energy along the path gives an estimate of the activation energy, an essential parameter for estimating the transition rate within the harmonic approximation to transition state theory.\cite{Wigner1938,Vineyard1957} It can also be used to parameterize a reversible work evaluation of the free energy barrier needed for the estimation of the transition rate in classical and quantum transition state theory.\cite{Schenter1994} Paths can be represented in a discrete way by a sequential set of atom configurations referred to as {\it images} of the system (see ref.~\cite{Asgeirsson2020} for a review), as in the nudged elastic band (NEB) method \cite{Mills1995,Jonsson1998} which is widely used for finding MEPs. NEB is a local optimization method starting from some initial path that should ideally represent a good guess. The computational efficiency of the method is in part due to the effective use of parallel computing, as the calculation of the energy and atomic forces on all the images representing the path can be carried out simultaneously. The success and computational efficiency of an NEB calculation, however, depends on the quality of the path from which the calculation starts. The closer the initial path is to the MEP, the smaller the computational effort will be to reach convergence. The quality of the initial path is particularly important when the energy and atomic forces are evaluated using an electronic structure method requiring significant computational effort. Furthermore, sometimes more than one MEP exists connecting given initial and final states, and since the NEB method represents a local optimization, it will converge on the one closest to the initial path. A good initial path is of particularly great importance in such cases.

The simplest way of generating an initial path is linear interpolation of the Cartesian coordinates of the initial and final state configurations. This method works well in many cases and has in particular become widely used in computational studies of heterogeneous catalysis as well as diffusion in and on the surfaces of solids.\cite{Asgeirsson2020} The linear interpolation can, however, lead to the creation of unrealistic atomic configurations in the intermediate images resulting in large atomic forces and even problems with convergence in the electronic structure calculations. Such problems can, for example, arise when significant rotation of a group of atoms occurs during the transition, as shown below. One way to address such problems is to define a minimum distance between atoms and/or construct a linear interpolation to and from an intermediate configuration that represents a good guess for a transition structure. This approach, however, makes the calculations more tedious and requires significant user input, which is highly undesirable in the context of large scale studies using MEP calculations. Internal coordinates can work better than Cartesian coordinates in part because of the inclusion of dihedral angles,\cite{Kastner2009} but the complexity of the implementation is then greatly increased, and there is still no guarantee that the initial path will be good since large rotating groups can still overlap when the linear interpolation is carried out. 

An alternative to the linear interpolation and subsequent optimization of the images in parallel is a sequential generation of the MEP starting from each of the two endpoints until the two path segments meet. This approach has been developed and used in electronic structure calculations of various reactions.\cite{Peters2004,Zimmerman2013_1,Zimmerman2013_2} By growing the path sequentially, unphysical high-energy configurations that may be generated in a linear interpolation can be avoided. This method, however, has the disadvantage that the efficient use of parallel computation in simultaneous calculation of all the images is no longer possible.

A more recent approach is to generate an initial path on the so-called image dependent pair potential (IDPP).\cite{Smidstrup2014} There, an interpolation of the distance between pairs of atoms in the initial and final state configurations is carried out for each of the intermediate images. A weighted sum over squared differences between the pairwise distances of a given path and the interpolated values gives an objective function surface for which an NEB calculation is carried out to get the path that matches the interpolated pairwise distances as closely as possible. In the previous implementation, the NEB calculation on the IDPP surface starts from a linear interpolation of the Cartesian coordinates of the endpoints. This approach will be referred to as the LI-IDPP method. In many cases, LI-IDPP has been shown to generate significantly better initial paths than a linear interpolation of the Cartesian coordinates of the endpoint configurations. However, large rotations of groups of atoms during the transition can cause the linear interpolation in the initial phase of the LI-IDPP method to generate intermediate configurations where atoms come too close to each other and unnecessarily break bonds. Several examples of such problems are given below. An NEB calculation starting from such an initial path will converge on an unphysical MEP with high activation energy. 

In this article, we present a variation on the use of the IDPP surface where the initial path is generated sequentially. This approach will be referred to as the S-IDPP method. The computational effort is minimal since the IDPP surface can be evaluated readily, as it does not involve an evaluation of the energy of the system. In this way, a linear interpolation of Cartesian coordinates is avoided altogether, and a better initial path is generated for a subsequent calculation of the MEP using the NEB method taking advantage of parallel computing. The article is structured as follows: In section~\ref{methodology}, the LI-IDPP method is briefly reviewed and the S-IDPP method presented with an illustration using a two-dimensional model surface. Section~\ref{settings} contains information about the electronic structure calculations that are presented as applications, and section~\ref{results} describes the results of S-IDPP calculations as well as a comparison with LI-IDPP calculations. A discussion of the results is given in section~\ref{discussion}, and conclusions are presented in section~\ref{conclusion}.
\section{Methodology}\label{methodology}
In this section, the notation is defined, the LI-IDPP reviewed briefly for completeness, and the S-IDPP method presented. Then, the computational methodology and parameters are given for the molecular reactions that are presented as applications of the S-IDPP method.

The requested number of images in the discrete representation of the path is denoted as $N_{\mathrm{im}}^{\mathrm{req}}$. 
As for any calculation based on discretization of a variable, it is important to use a large enough number of discretization points. Since NEB calculations are typically carried out in combination with computationally intensive electronic structure calculations, the number of images is usually kept to a minimum, typically between 5 and 10. However, when working with the IDPP surface, the calculations are fast and it can be advantageous to use a larger number of images when the path is generated, for example twice the number of images than eventually will be used in the NEB calculations. Most importantly, a larger number of images gives a better estimate of the local tangent. In the subsequent MEP calculations, the number of images in the converged S-IDPP path can be reduced, for example by including only every other image.

The Cartesian coordinates of the images are $\mathbf{r} = \left\{r_1, r_2, \dots , r_{N_{\mathrm{im}}^{\mathrm{req}} - 1}, r_{N_{\mathrm{im}}^{\mathrm{req}}} \right\}$\,, where $r_1$ and $r_{N_{\mathrm{im}}^{\mathrm{req}}}$ are the fixed atom coordinates corresponding to the initial and final configurations of the atoms. These quantities are vectors with 3$N_{\mathrm{at}}$ coordinates where $N_{\mathrm{at}}$ is the number of atoms in the system. The three Cartesian coordinates of atom $j$ in image $l$ are denoted by $r_{l,j}$. 

Before the calculation starts, it is important to shift and rotate the initial and final state configurations so as to minimize the root-mean-square deviation between the structures by using, for example, quaternion rotation\cite{Coutsias2004,Melander2015} or Kabsch alignment.\cite{Kabsch1976}. Otherwise, the path becomes unnecessarily long because of irrelevant translation and/or rotation.
\subsection{Image dependent pair potential}
The IDPP is based on an interpolation of the distances between pairs of atoms in the initial and final states. For a path with equal spacing of images, the interpolated distance between atoms $i$ and $j$ corresponding to image $l$ is
\begin{equation}
d_{ij}^{l} = d_{ij}^{R} + \frac{l - 1}{N_{\mathrm{im}}^{\mathrm{req}} - 1} \, \left(d_{ij}^{P} - d_{ij}^{R}\right)\,,
\end{equation}
where $d_{ij}^{R}=\left|{r_{1,j}-r_{1,i}}\right|$ is the distance between the atoms in the initial state configuration, and the expression is analogous for the final state, $d_{ij}^{P}$. For a given path, an objective function denoting how well image $l$ matches the interpolated values is given by the weighted sum of squared differences between the atom distances in the image and the interpolated values  
\begin{equation}
S_{l}^{\mathrm{IDPP}}\left(r_{l}\right) = \sum_{i = 1}^{N}\sum_{j > i}^{N}\omega \left(\left|r_{l,j}-r_{l,i}\right|\right)\left(\left|r_{l,j} - r_{l,i}\right| - d_{ij}^{l}\right)^{2}\,.
\end{equation}
A weight function $\omega \left(x\right) = x^{-4}$ is typically used to place greater emphasis on short atom distances. This IDPP objective function can be used instead of an energy surface in an NEB calculation to give a path that is locally optimal in a least squares sense. 

The NEB method involves defining an effective force on the atoms in such a way as to ensure convergence to an optimal path on the objective function surface and giving control over the location of the images along the path when the forces are zeroed using some optimization algorithm. The effective force consists of the component of the gradient of the objective function (usually the energy of the system, but here the IDPP objective function) perpendicular to the path and a spring force that controls the distribution of the images along the path. The NEB force acting on the atoms in image $l$ is 
\begin{equation}
F_{l}^{\mathrm{NEB}} = F_{l}^\perp + F_l^{\parallel\hspace{1pt}\mathrm{sp}}  
\end{equation}
with
\begin{equation}
F_{l}^\perp = -{\nabla S_{l}^{\mathrm{IDPP}}} + \left(\nabla S_{l}^{\mathrm{IDPP}} \cdot \hat{\tau}_l\right)\hat{\tau}_{l}
\end{equation}
and
\begin{equation}
F_{l}^{\parallel\hspace{1pt}\mathrm{sp}} = \left(k_{l}\left|r_{l + 1} - r_{l}\right| - k_{l - 1}\left|r_{l} - r_{l - 1}\right|\right)\hat{\tau}_{l} \,,
\end{equation}
where $\hat{\tau}_l$ is a normalised estimate of the tangent to the path at image $l$. The spring constant $k_{l}$ is for line-segment $\left[r_l, r_{l + 1}\right]$. If all the spring constants are chosen to have the same value, equal spacing of the images will result. In the application below, the distance between images is necessarily quite different during the sequential construction of the path and it is, therefore, important to choose unequal values of the spring constants. In regular NEB calculations, it can also be advantageous to choose unequal spring constants, for example weighted by the energy, in order to get higher density of images near the top of an energy barrier.\cite{Henkelman2000_1, Asgeirsson2020} 

The tangent to the path at image $l$ is estimated by the line segment to the neighboring image with higher IDPP objective function value (the superscript "IDPP" in $S^{\mathrm{IDPP}}$ is omitted for brevity)
\begin{equation}
{\tau}_{l} = \begin{cases}
				{\tau}_{l}^{+} = r_{l + 1} - r_{l}\,,\ \ \  \mathrm{ if } \  S_{l+1} > S_{l} > S_{l-1}\\
				{\tau}_{l}^{-} = r_{l} - r_{l - 1}\,, \ \ \  \mathrm{ if } \  S_{l+1} <S_{l} < S_{l-1}
			\end{cases}\hspace{-10pt},
\end{equation}
and a weighted average is used for an image that has the locally highest or lowest value of the objective function
\begin{equation}
    \label{eq:growing-neb-tangent-extrema-and-frontier}
{\tau}_{l} = \begin{cases}
	\hat{\tau}_{l}^{+}\Delta S_{l}^{\mathrm{max}} + \hat{\tau}_{l}^{-}\Delta S_{l}^{\mathrm{min}}\,,\ \ \  \mathrm{ if }\  S_{l + 1} > S_{l - 1}\\
	\hat{\tau}_{l}^{+}\Delta S_{l}^{\mathrm{min}} + \hat{\tau}_{l}^{-}\Delta S_{l}^{\mathrm{max}}\,, \ \ \  \mathrm{ if }\  S_{l + 1} < S_{l - 1}
			\end{cases}\hspace{-10pt},
\end{equation}
where
\begin{equation}
\Delta S_{l}^{\mathrm{max}} = \mathrm{max}\left(\left|S_{l + 1} - S_{l}\right|, \left|S_{l - 1} - S_{l}\right|\right)
\end{equation}
and
\begin{equation}
\Delta S_{l}^{\mathrm{min}} = \mathrm{min}\left(\left|S_{l + 1} - S_{l}\right|, \left|S_{l - 1} - S_{l}\right|\right)\,.
\end{equation}
Note that a normalised tangent is used here, given by $\hat{\tau}_{l} = {\tau}_{l}/|{\tau}_{l}|$. The difference between eq.~\ref{eq:growing-neb-tangent-extrema-and-frontier} and the  tangent definition in ref.\,\cite{Henkelman2000_2} is that ${\tau}_{l}^{+}$ and ${\tau}_{l}^{-}$ are normalised before the calculation of the weighted average. This modification is found to be necessary for the S-IDPP method.

As for any NEB calculation, an initial path is needed as a starting point for the optimization using the IDPP objective function. In the original IDPP method, LI-IDPP, a linear interpolation of the Cartesian coordinates of the endpoints is used as the initial path. As will be shown in several examples below, the local optimization starting from such a path can lead to problematic solutions where bonds are unnecessarily broken and then reformed along the path. The problem lies in using the linear interpolation of Cartesian coordinates as an initial path. 

The optimization of image $l$ involves shifting the location of the atoms in the direction of the effective force, ${ F_l}^{\mathrm{NEB}}$, until the force has been zeroed. For this purpose, the velocity projection optimization algorithm\cite{Jonsson1998,Asgeirsson2020} is applied using a small enough time step to avoid irregularities in the path.
\subsection{S-IDPP}\label{methodology-gr-neb}
Instead of the initial linear interpolation of the reactant and product atomic configurations, the S-IDPP method constructs the path sequentially starting from the endpoints. The first images are added close to the endpoint configurations, then images farther away, etc.,  until the path contains the requested number of images, ${N_{\mathrm{im}}^{\mathrm{req}}}$. At each stage in the sequential construction of the path, spring constants are chosen to give roughly equal spacing of the images in each of the two parts of the path near the endpoints, but larger spacing between the two parts. Finally, an NEB calculation on the IDPP surface is carried out to refine the path, mainly to adjust the distribution of the images along the path.

The first iteration of the path involves adding two images, one at $r_2$ near the reactant and another one at $r_{N_{\mathrm{im}}^{\mathrm{req}} - 1}$ near the product configuration. The corresponding spring constants are chosen to take the large difference in the distance between these images into account as compared to their distance from the nearest endpoint. When $k$ denotes a typical value for a spring constant when equal spacing of the images is desired, the spring constant for these newly added images is chosen to be 
\begin{equation}
k_{2} = \frac{d_{\mathrm{id}}}{\left|r_{2} - r_{N_{\mathrm{im}}^{\mathrm{req}} - 1}\right|}\,k\,,
\end{equation}
where $d_{\mathrm{id}}$ is the ideal distance between images if $N_{\mathrm{im}}^{\mathrm{req}}$ images are distributed evenly along the current path, which at this point is just a linear interpolation between the endpoints
\begin{equation}
d_{\mathrm{id}} = \frac{\left|r_{1} -  r_{N_{\mathrm{im}}^{\mathrm{req}}}\right|}{\left(N_{\mathrm{im}}^{\mathrm{req}} - 1\right)}\,.
\end{equation}
The positions of the images are adjusted in order to zero the NEB force to a given tolerance. Because of the choice of spring constants, the newly added images are close to the endpoints, but they are separated from each other by a large distance.

As soon as the optimization of one of the images has reached convergence, a new image is added to that part of the path, initially placed a distance $d_{\mathrm{id}}$ from the previously placed image, in a direction given by the tangent (always eq.~\ref{eq:growing-neb-tangent-extrema-and-frontier}). At each introduction of a new image, the total length of the current path is updated, simply by summing over the length of the linear segments, the ideal distance between images reevaluated, and values for the spring constants chosen in order to place the image at an appropriate distance from the previous one. The NEB force is zeroed by adjusting the positions of all images currently placed on the path. This process is repeated until the requested number of images has been placed.

A flowchart of the algorithm is shown in figure 1 and an example calculation shown in figure 2 for different stages of the sequential construction of the path.  This example shows how the path has the tendency to follow the slowest ascent direction which in this case is different from the optimal path. However, eventually, as more images are added (beyond 9 in this case), the path recovers and converges on the optimal path.
\begin{figure}
	\includegraphics[width = \textwidth]{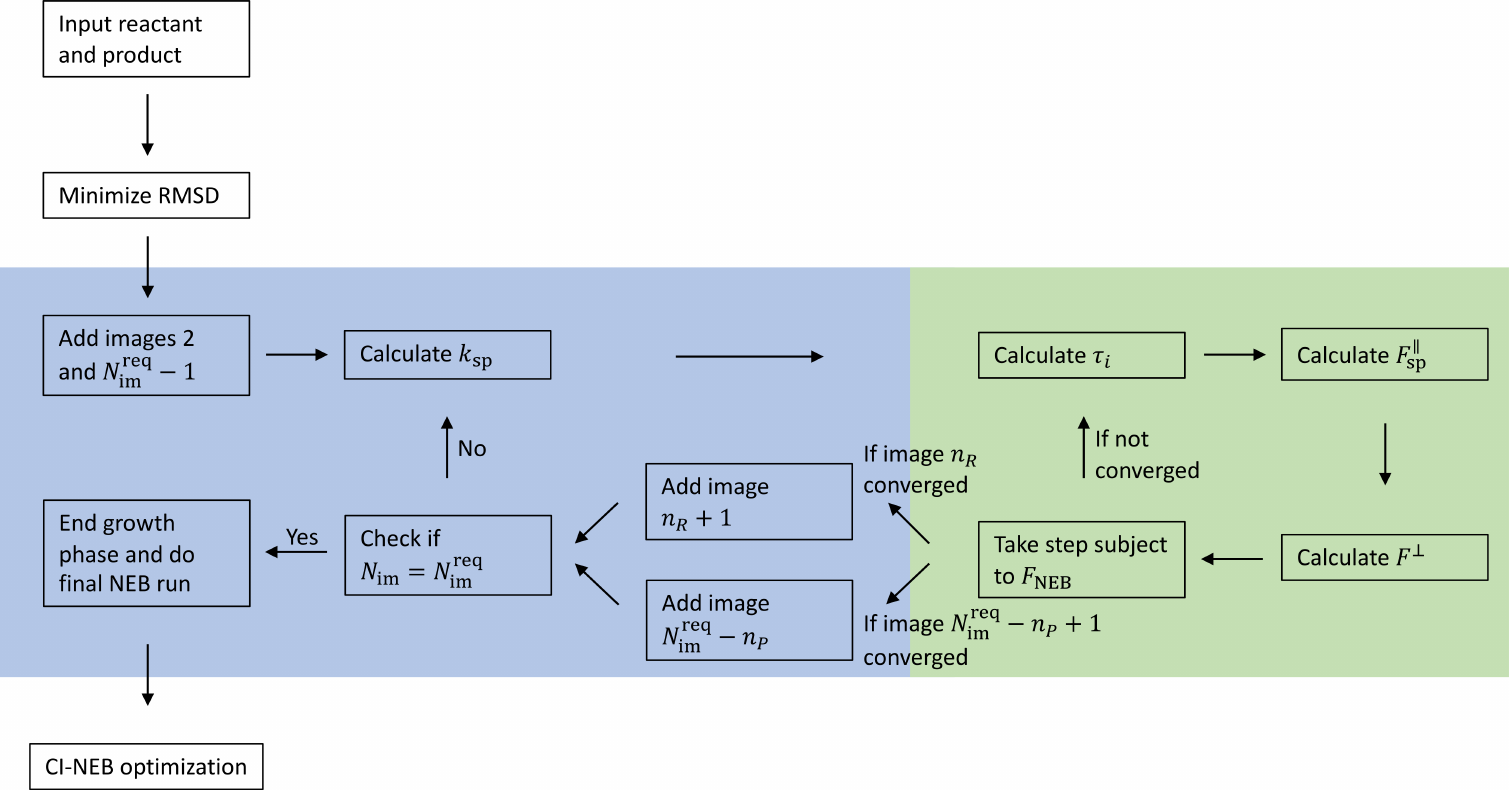}
	\caption{Flowchart of the S-IDPP method. The steps where images are added are shown with a blue background, while the optimization steps are shown with a green background. The current number of images is denoted by $N_{\mathrm{im}}$, while $N_{\mathrm{im}}^{\mathrm{req}}$ is the requested total number of images. The current number of images in the parts of the path near the reactant and product configurations are denoted by $n_{R}$ and $n_{P}$, respectively. Once the number of images reaches the requested number, an NEB calculation on the IDPP surface is carried out with equal spring constants to refine the path, especially the distribution of images along the path. The resulting path is then ready for use in a CI-NEB calculation using an electronic structure method to estimate the energy and atomic forces.}
	\label{fig:growing-neb-flowchart}
\end{figure}

Once the requested number of images has been added, an NEB calculation is carried out with equal spring constants to refine the path and distribute the images evenly along the path. Then, the path is ready as an initial path for an MEP calculation using the CI-NEB method\cite{Henkelman2000_1} in combination with some method for estimating the energy and atomic forces, typically an electronic structure calculation, as illustrated in the examples given below. There, it can be advantageous to use energy weighted springs.\cite{Henkelman2000_1,Asgeirsson2020} 
\begin{figure}
	\includegraphics{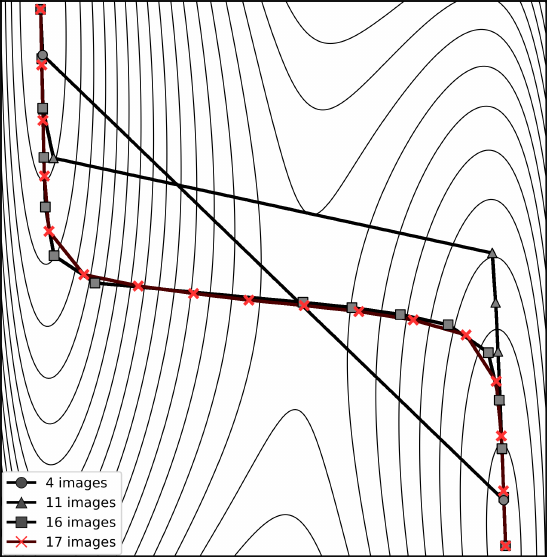}
	\caption{Illustration of the S-IDPP method for a 2-dimensional model IDPP surface. The path is visualised for intermediate steps after adding 2 images (circles), 9 images (triangles) and 14 images (squares) using the spring constant scaling algorithm. An optimal path obtained in an NEB calculation with 17 images is shown by red crosses. Note how the images initially line up along the slowest ascent path from the lower right endpoint (triangles), but later converge to the optimal path.}
	\label{fig:lepsho-growing-neb-growth}
\end{figure}
\subsection{Computational Settings}\label{settings}
The convergence thresholds used in the S-IDPP calculations are $0.01\,\mathrm{Å}^{-3}$ for the maximum atomic force component and $0.005\,\mathrm{Å}^{-3}$ for the RMS value. The base spring constant is 1\,$\mathrm{Å}^{-2}$. 
The chosen number of images in the S-IDPP construction matches the one used in the subsequent CI-NEB calculation in all cases except the rotation of 9,9'-bianthracene where using twice as many images in the S-IDPP calculation is found to improve the quality of the initial path.

The subsequent CI-NEB calculation using electronic structure estimates of energy and atomic forces in the molecular reactions make use of 9 images to describe the path. The B3LYP+D3(BJ)/def2-SVP\cite{Lee1988, Becke1988, Becke1993, Grimme2006, Weigend2005} and GFN2-XTB\,\cite{Grimme2017} calculations are performed with the ORCA software\cite{Neese2012, Neese2022} using energy-weighted spring constants between $0.01$ and $0.1$\,a.u.\ and are converged to a threshold of $2.5\cdot 10^{-3}$\,a.u.\ for the RMS of the force orthogonal to the image tangent and $5\cdot 10^{-3}$\,a.u.\ for the largest perpendicular force component for all images. The total force on the climbing image is converged to threshold values one order of magnitude tighter than for the other images.


\section{Results}\label{results}
We present four examples of molecular reactions where the LI-IDPP method turns out to give unphysical paths that are so different from the optimal MEP that subsequent NEB calculations would converge on an MEP with very high activation energy. The S-IDPP method, however, gives initial paths that are realistic, and the subsequent NEB calculations using electronic structure calculations to estimate energy and atomic forces converge on what appear to be optimal MEPs. The first system is the Diels-Alder reaction between ethylene and cyclopentadiene. The second reaction is an isomerisation of tris[(3-methyl-1H-benzimidazol-1-yl-2(3H)-ylidene)-1,2-phenylene]iridium(III) (TMBPI) due to the rotation of one of its [(3-methyl-1H-benzimidazol-1-yl-2(3H)-ylidene)-1,2-phenylene]\textsuperscript{-} (MBP) ligands by 180°. The third reaction is a [3+2]-cycloaddition reaction of 9-(2-azidoethyl)-10-(but-3-yn-1-yl)anthracene, in which an azidoethyl group rotates around a large aromatic group. The fourth reaction involves sterically hindered rotation of an anthryl group in 9,9'-bianthracene.


\subsubsection{Diels-Alder Reaction}\label{diels-alder}
\begin{figure}
	\includegraphics[width = \textwidth]{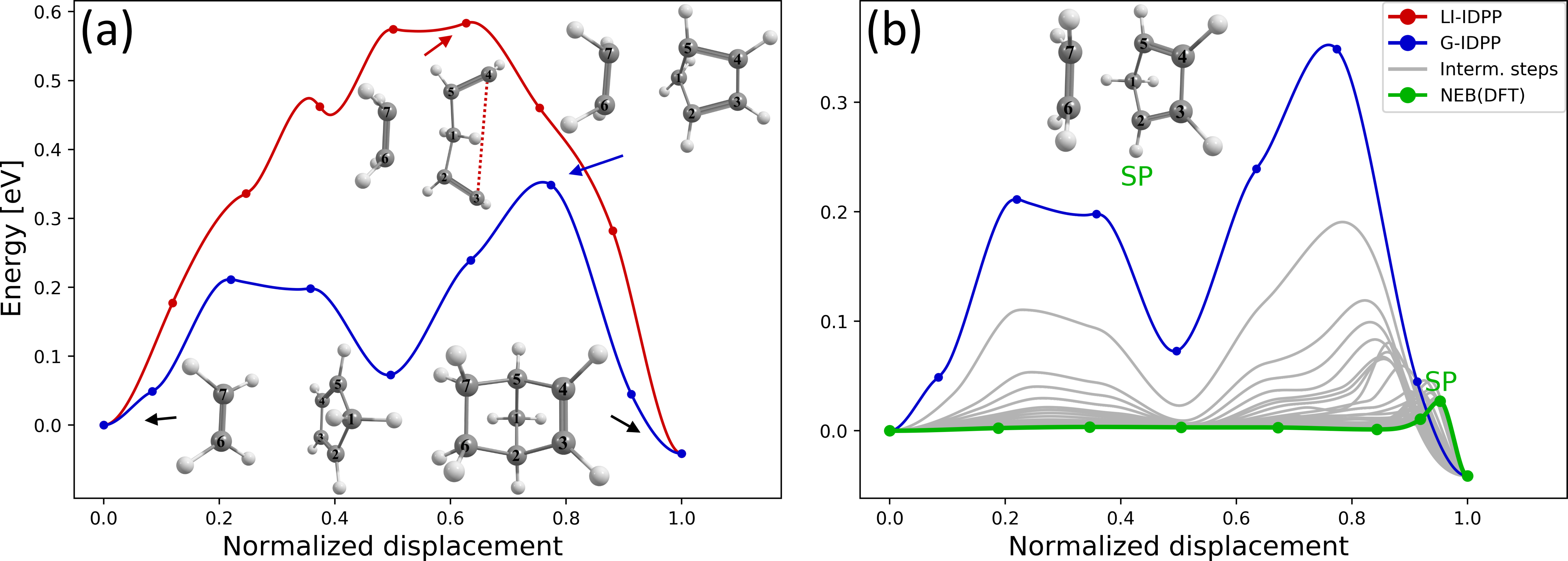}
	\caption{Diels-Alder reaction of ethylene and cyclopentadiene. 
    (a) Energy profiles calculated using B3LYP+D3(BJ)/def2-SVP along paths obtained by LI-IDPP (red) and S-IDPP (blue). The reactant, product and highest-energy image configurations are shown as insets. Along the LI-IDPP path, 
    a C-C bond is unnecessarily broken, as highlighted by the red dashed line. In contrast, the S-IDPP path preserves the structural integrity of the cyclopentadiene molecule. 
    (b) Energy along the path for various stages of the CI-NEB calculation starting with the S-IDPP path (blue as in (a)). Intermediate stages of the optimization (gray) and the final converged MEP (green) are shown. The saddle point configuration corresponding to the maximum along the MEP is included as an inset.
    }
	\label{fig:idpp-paths-resolved-8}
\end{figure}

The reactant and product configurations of this Diels-Alder reaction system have been taken from ref.~\citenum{Zimmerman2013_1}. The molecules in this system are unfavorably oriented with respect to each other. The necessary rotation of the cyclopentadiene fragment in the initial structure of the system poses problems for the LI-IDPP method, as is visualized in fig.~\ref{fig:idpp-paths-resolved-8}. The bond between carbon atoms 3 and 4 breaks and reforms over the course of the path. The S-IDPP method produces a significantly better path preserving the structural integrity of the cyclopentadiene molecule. The two paths represent different locally optimal paths on the IDPP surface that differ because of the different ways the paths are constructed.

The subsequent CI-NEB calculation using the energy and atomic forces estimated by the B3LYP+D3(BJ)/def2-SVP electronic structure method and the S-IDPP initial path converges to a sensible MEP that has low activation energy. The long tail of the MEP corresponds to a 180$^\circ$ rotation of the cyclopentadiene fragment. 


\subsubsection{Isomerisation Reaction}\label{isomerisation}
\begin{figure}
	\includegraphics[width = \textwidth]{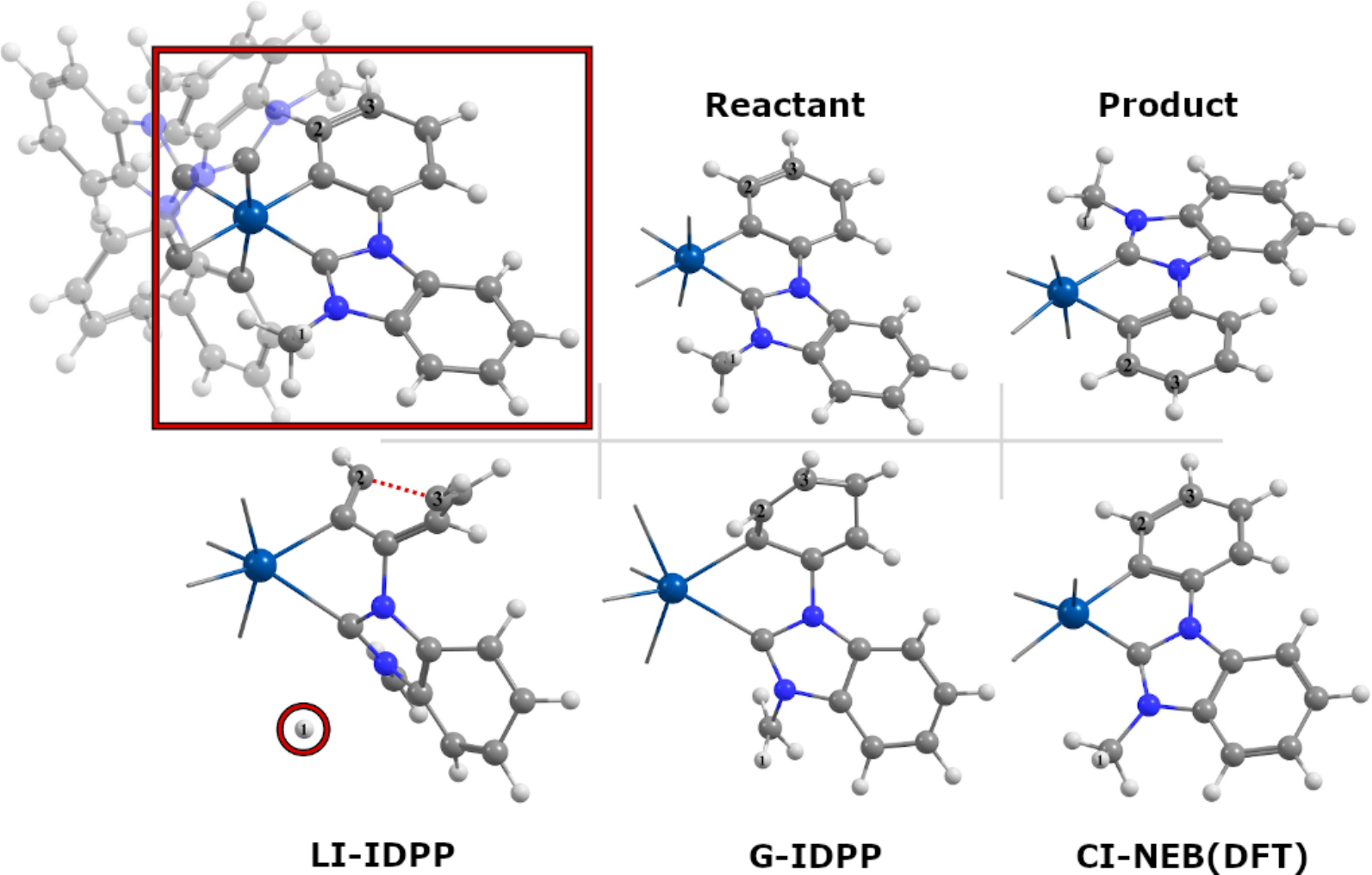}
	\caption{Isomerization of TMBPI where the MBP ligand rotates by 180$^\circ$. The full Ir-complex (left) and the most relevant region of the reactant (center) and product (right) configurations are shown in the upper panel. The lower panel shows the image with highest energy when the LI-IDPP path (left) and S-IDPP path (center) are used in a GFN2-XTB calculation without further optimization, and then the saddle point configuration (right) obtained by a CI-NEB calculation started from the S-IDPP path, using first GFN2-XTB and subsequently a saddle point optimization at the B3LYP+D3(BJ)/def2-SVP level. The red circle and the dashed line in the highest image of LI-IDPP highlight the problems occurring there, where a C-C bond is broken and an H atom dissociates from the ligand. The S-IDPP path, however, yields a highest-energy configuration that is quite similar to the one obtained in the subsequent DFT calculation.}
	\label{fig:idpp-paths-cr}
\end{figure}

The isomerisation reaction between the facial and meridional isomers of metal carbene complexes is of interest since these complexes offer favorable properties for diverse applications.\cite{Osiak2017} The properties of the two isomers differ significantly. Here, the isomerisation of TMBPI in the absence of an acid is calculated. This reaction  is challenging because of the large rotation involved in the isomerisation reaction. As shown in fig.~\ref{fig:idpp-paths-cr}, a large group of atoms constituting an MBP ligand rotates by about 180°. The LI-IDPP method produces a path where the structural integrity of this rotating ligand is lost, as the bond between carbon atoms 2 and 3 is broken and then later reformed over the course of the reaction. Additionally, hydrogen atom 1 is dissociated from the molecule. As a result, the LI-IDPP path is unsuitable as an initial path for a CI-NEB calculation in an electronic structure calculation of the MEP. The S-IDPP path, however, preserves the structural integrity of the ligand and gives a path that can be used as an initial path in a CI-NEB calculation with an electronic structure method. First, the MEP is found on the GFN2-XTB energy surface. Then, the saddle point is determined more accurately using the final position of the climbing image in the GFN2-XTB calculation as input and carrying out a saddle point optimization at the B3LYP+D3(BJ)/def2-SVP level. The resulting saddle point configuration is shown in fig.~\ref{fig:idpp-paths-cr}. The highest-energy image on the S-IDPP path is quite similar showing that the S-IDPP path is a good initial guess for the electronic structure calculation, while the LI-IDPP path is not.


\subsubsection{[3+2]-Cycloaddition Reaction}\label{click}
\begin{figure}
	\includegraphics[width = \textwidth]{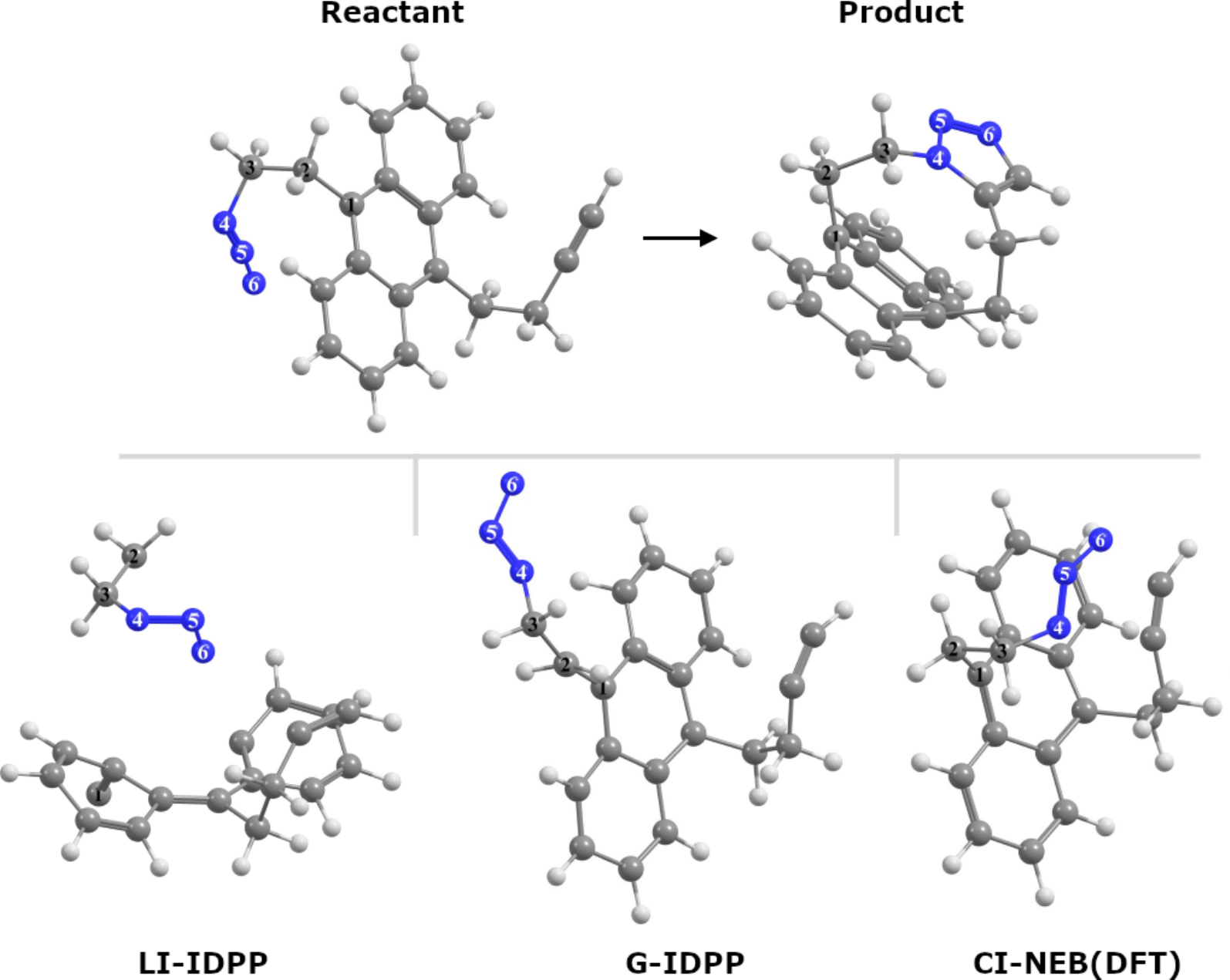}
	\caption{
	[3+2]-cycloaddition reaction of 9-(2-azidoethyl)-10-(but-3-yn-1-yl)anthracene. In the upper panel, the reactant (left) and product (right) configurations are shown. In the lower panel, the highest-energy image configurations along the LI-IDPP (left) and S-IDPP (center) paths are shown along with the saddle point configuration (right) obtained by a CI-NEB calculation using B3LYP+D3(BJ)/def2-SVP starting from the S-IDPP path. In the image from the LI-IDPP path, the azidoethyl group is dissociated from the molecule making this path a poor initial guess for the electronic structure calculation of the MEP.}
	\label{fig:idpp-paths-click}
\end{figure}

Figure~\ref{fig:idpp-paths-click} shows a [3+2]-cycloaddition of 9-(2-azidoethyl)-10-(but-3-yn-1-yl)anthracene. The main challenge here for the initial path generation is not the cycloaddition itself, but rather the rotation of the azidoethyl group to the other side of the aromatic rings. In the LI-IDPP path, the azidoethyl group is dissociated from the molecule, and a dangling bond is formed on carbon atom 1. In the S-IDPP path, however, the azidoethyl group rotates to the other side of the aromatic groups without any bond breaking, and the highest-energy image corresponds to the reactive step of the cycloaddition, as in the MEP on the B3LYP+D3/def2-SVP surface obtained with the subsequent CI-NEB calculation.


\subsubsection{Rotational Isomerization}\label{rotation}
\begin{figure}
	\includegraphics[width = \textwidth]{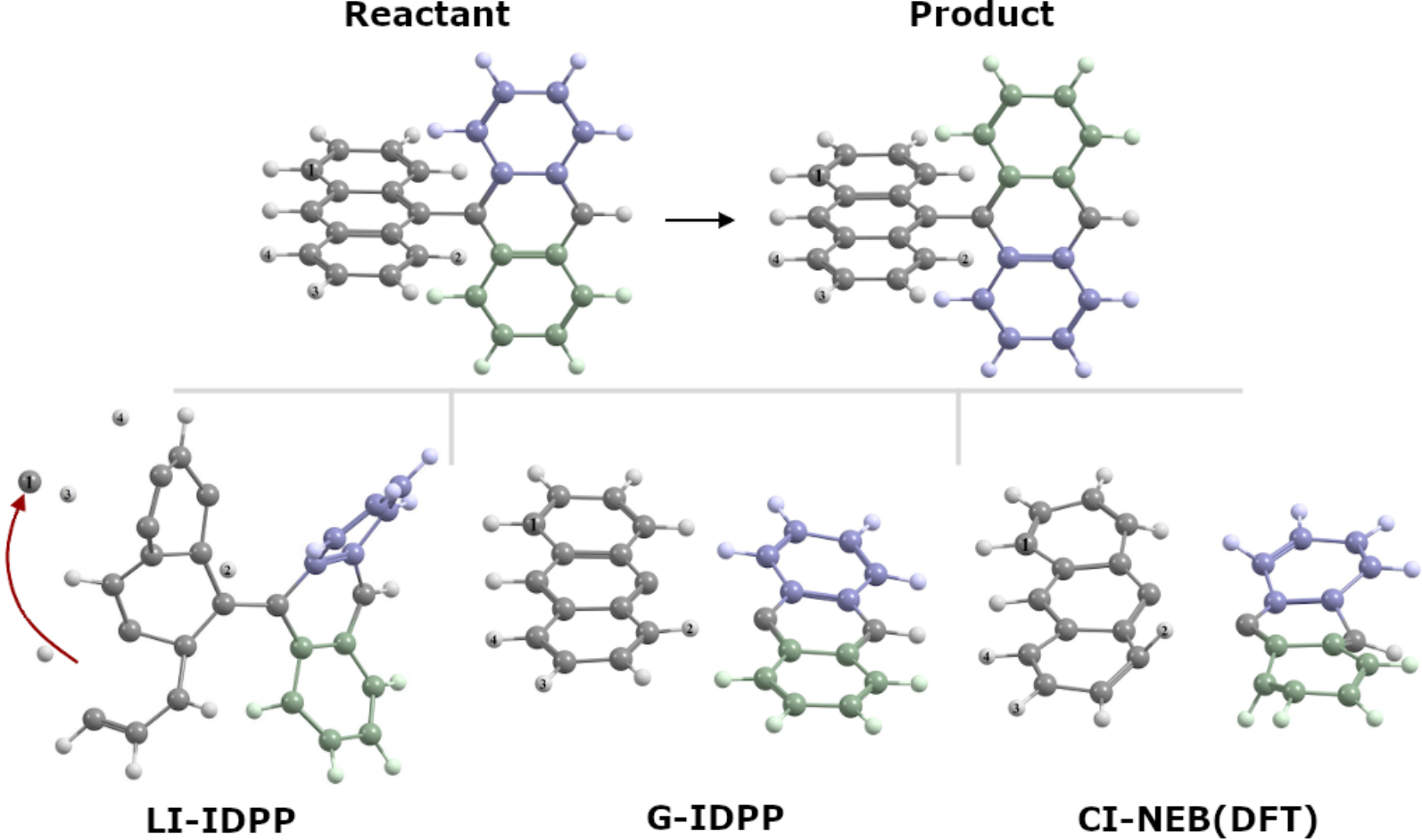}
	\caption{
	Rotation of an anthryl group in 9,9'-bianthracene. In the upper panel, the reactant (left) and product (right) configurations of the reaction are shown. In the lower panel, the highest-energy image along the LI-IDPP (left) and S-IDPP (center) paths obtained using B3LYP+D3(BJ)/def2-SVP without further relaxation are shown along with the saddle point configuration (right) obtained from a converged CI-NEB calculation started from the S-IDPP path. The rotating anthryl group is colored for clarity. In the LI-IDPP path, a C atom and multiple H atoms dissociate from the molecule. The red arrow points to a C atom that is abstracted from one of the aromatic rings.}
	\label{fig:idpp-paths-bianth}
\end{figure}

The rotation of an anthryl group in 9,9'-bianthracene is shown in fig.~\ref{fig:idpp-paths-bianth}. One of the two anthryl groups rotates by 180°. This rotation is sterically hindered enough that the $\sigma$-bond around which the rotation occurs breaks and is then reformed subsequently along the MEP calculated at the B3LYP+D3(BJ)/def2-SVP level. In the LI-IDPP path, a carbon atom and multiple hydrogen atoms dissociate from the anthryl group that does not rotate. In addition, the planarity of both of the anthryl groups is lost. However, the S-IDPP path maintains structural integrity in both anthryl groups, and the distortion of the aromatic groups is much less pronounced, leading to a correct description of the rotation. The highest-energy image of the S-IDPP path is quite similar to the highest-energy image along the MEP on the B3LYP+D3(BJ)/def2-SVP surface. 
For this reaction, it turns out to be better to use a larger number of images in the S-IDPP calculation, 17 rather than 9.  The subsequent MEP computation at the B3LYP+D3(BJ)/def2-SVP level is started with only every other image of the converged S-IDPP path. Such an increase in the number of images does not, however, improve the path obtained with the LI-IDPP method.


\section{Discussion}\label{discussion}
The central problem with the LI-IDPP method is the linear interpolation of Cartesian coordinates between the endpoint structures prior to the path optimization on the IDPP surface. For all the problematic reactions presented here, this interpolation places some atoms too close to each other, typically around the midpoint of the initial path. At the point along the path at which the distances become minimal, these atoms are approximately in the same plane which we refer to as a partitioning plane in the linear interpolation path. The IDPP surface was formulated in order to increase too small atom distances and thereby improve upon the linear interpolation originally used in NEB calculations. For atoms close to a partitioning plane, the weight function contribution, which only depends on the atom distances without taking chemical bonding into account, dominates the IDPP force by far and points away from the partitioning plane. As atoms move through a partitioning plane, the direction of the full IDPP gradient, in particular its perpendicular projection on the path tangent, can change dramatically, even up to a complete reversal. If one part of a group of atoms has moved past a partitioning plane, while another has not, the perpendicular IDPP gradient will point away from the partitioning plane in opposite directions for the two parts of the group leading to fragmentation during the IDPP path optimization, even if conservative settings are used for the optimization. The isomerisations of TMBPI and 9,9'-bianthracene are examples of this effect.

The accumulation of atoms in a partitioning plane can also lead to atoms being placed inside bonds between other atoms, in which case the bond is broken during the path optimization on the IDPP. The Diels-Alder reaction is a testament to this effect. In the [3+2]-cycloaddition, a partitioning plane is formed by the anthracene aromatic plane and already visible in the endpoint structures. In this case, atoms are placed inside the aromatic plane in intermediate images, destroying the structural integrity of the plane during the IDPP NEB optimization.

When an odd number of images is used, these problems with the linear interpolation are more likely to be encountered since an image is then placed at the midpoint of the path where a partitioning plane is most likely to occur. By choosing an even number of images, the problems may be avoided, but not in all cases, in particular not for the four systems studied in this article. Once structural integrity is lost in one image along the path, the parallel spring forces between the images gradually disperse the problem along the entire path due to pairwise coupling of adjacent images. The S-IDPP method presented here avoids these problems by not starting from a linear interpolation of the Cartesian coordinates.

In particularly challenging cases, such as the rotation of 9,9'-biantracene, the quality of the S-IDPP initial path can be improved by using more images for the S-IDPP calculation than for the subsequent MEP calculation. The reason for this improvement is the decrease in optimal spacing between the images when more images are used. As a result, new images are placed at shorter distance from converged images meaning they venture less into the unknown region of the IDPP surface. The same cannot be said for LI-IDPP since there, all images are placed on the path immediately, which leaves little opportunity for improvement when using more images. The simplest way to reduce the number of images between the two paths is to use twice the number of images (one less for an odd number of images) for the initial path and then using every second image for the MEP calculation.

It is worth noting that some of the problems with the linear interpolation could be solved manually. The unfavorable alignment of fragments in the Diels-Alder reaction can be resolved by changing one of the endpoint structures. A useful initial path for the isomerisation of TMBPI can be constructed with LI-IDPP if a good guess for the saddle point configuration is used as an intermediate image along the path and two linear interpolations constructed, one from the initial configuration to this guess for the saddle point, and the other from there to the endpoint configuration. However, these strategies require human intervention and introduce bias into the calculation, so they are not satisfactory solutions, especially in the context of large scale studies using NEB calculations. The S-IDPP method constitutes a way of solving these problems automatically and without introducing bias into the computation.


\section{Conclusion}\label{conclusion}
The S-IDPP method is an improvement on the LI-IDPP method used to  generate initial paths for MEP calculations. 
Problems with the LI-IDPP method derive from the use of an initial interpolation of Cartesian coordinates at the start of the calculation. Atoms can then move through a partitioning plane resulting in an abrupt change of the direction of the perpendicular IDPP force acting on the atoms. The root of the problem is not numerical instability, which could be eliminated by reducing the step size of the optimizer, but rather the dominance of the weight function in the IDPP surface for atoms near the partitioning plane. S-IDPP avoids most of the linear interpolation of Cartesian coordinates and therefore, represents a significant improvement over LI-IDPP.
A two-dimensional test surface is used to demonstrate the concept of the S-IDPP method. It demonstrates that even when the slowest ascent direction differs from that of the optimal path, convergence to a good initial path can be obtained.


\begin{acknowledgement}
Y.L.A.S. thanks the Erasmus+ program of the European Union for a mobility grant. V.Á. acknowledges a doctoral fellowship from the University of Iceland. This work was supported by the Icelandic Research Fund and by the European Union's Horizon 2020 research and innovation program under grant agreement No 814416. The authors thank Prof.\ Bernd Engels for useful discussions. The calculations were carried out at the Icelandic High Performance Computing Center (IHPC).
\end{acknowledgement}


\begin{suppinfo}

The authors confirm that the data supporting the findings of this study are available within the article and/or its supplementary materials.

Data related to the results presented in this article is available at Zenodo.\cite{si_files}
Atomic configurations of all systems: Reactant and product configurations for Diels-Alder reaction of ethylene and cyclopentadiene; Reactant and product configurations for isomerisation of TMBPI; Reactant and product configurations for [3+2]-cycloaddition reaction of 9-(2-azidoethyl)-10-(but-3-yn-1-yl)anthracene; Reactant and product configurations for rotational isomerisation of 9,9'-bianthracene.

The S-IDPP method has been implemented in a branch of the EON software\cite{EON} that is part of the ReaxPro multiscale heterogeneous catalysis software suite.
\end{suppinfo}


\bibliography{Manuscript}

\newpage

\section{Atomic configurations of all systems}
All geometry parameters are given in \AA.
\subsection{Diels-Alder reaction of ethylene and cyclopentadiene}
\subsubsection{Reactant configuration}
\begin{table}[!h]
    \centering
    \begin{tabular}{ c c c c }
	H & -1.07081 & -2.21835 & 0.68124 \\
	H & 1.74670 & 1.26854 & -0.43957 \\
	H & 3.40352 & -1.29647 & 0.09967 \\
	H & -1.46930 & 1.36543 & -1.77800 \\
	H & -0.93779 & 2.21928 & 0.67592 \\
	H & 3.51034 & 1.18505 & 0.16951 \\
	H & 1.63908 & -1.19534 & -0.50456 \\
	H & -1.54919 & -1.34021 & -1.77498 \\
	H & 0.13419 & -0.03191 & 1.71674 \\
	H & -1.58776 & 0.02003 & 2.11400 \\
	C & 2.54849 & -0.67576 & -0.18815 \\
	C & -1.10667 & -1.17743 & 0.35768 \\
	C & -1.35015 & -0.72623 & -0.89481 \\
	C & -1.30651 & 0.74268 & -0.89659 \\
	C & -1.03610 & 1.18164 & 0.35475 \\
	C & 2.60638 & 0.65647 & -0.15090 \\
	C & -0.87608 & -0.00256 & 1.26677
    \end{tabular}
\end{table}
\FloatBarrier
\subsubsection{Product configuration}
\begin{table}[!h]
    \centering
    \begin{tabular}{ c c c c }
H & -0.11792 & 2.16141 & 0.69058 \\
H & 2.08994 & -1.18017 & -0.02456 \\
H & 1.14290 & 1.20885 & -1.52535 \\
H & -1.92208 & -1.33305 & -1.08980 \\
H & -0.11797 & -2.16141 & 0.69058 \\
H & 1.14287 & -1.20887 & -1.52535 \\
H & 2.08996 & 1.18012 & -0.02456 \\
H & -1.92205 & 1.33310 & -1.08979 \\
H & -0.91830 & 0.00001 & 2.04334 \\
H & 0.88169 & -0.00001 & 1.98245 \\
C & 1.18922 & 0.77880 & -0.51486 \\
C & -0.08626 & 1.12671 & 0.32357 \\
C & -1.27598 & 0.67156 & -0.50884 \\
C & -1.27600 & -0.67153 & -0.50884 \\
C & -0.08628 & -1.12671 & 0.32357 \\
C & 1.18920 & -0.77883 & -0.51486 \\
C & -0.04120 & 0.00000 & 1.37936
    \end{tabular}
\end{table}
\FloatBarrier
\subsection{Isomerisation of TMBPI}
\subsubsection{Reactant configuration}
\begin{table}[!h]
    \centering
    \begin{tabular}{ c c c c }
C & -3.28867 & 0.80367 & 1.53343 \\
N & -3.07193 & -0.40854 & 0.77019 \\
C & -1.92858 & -0.74766 & 0.11805 \\
N & -2.17065 & -1.96912 & -0.47579 \\
C & -3.47606 & -2.39512 & -0.21355 \\
C & -4.04664 & -1.38657 & 0.59613 \\
C & -4.24172 & -3.50778 & -0.57868 \\
C & -5.55642 & -3.58147 & -0.10718 \\
C & -6.10807 & -2.57908 & 0.70606 \\
C & -5.35555 & -1.45890 & 1.06878 \\
C & -1.10936 & -2.53778 & -1.24221 \\
C & 0.03914 & -1.71319 & -1.31043 \\
C & 1.11553 & -2.21951 & -2.05710 \\
C & 1.06931 & -3.47598 & -2.67448 \\
C & -0.07378 & -4.27000 & -2.55951 \\
C & -1.17682 & -3.79942 & -1.83837 \\
Ir & -0.11451 & 0.12020 & -0.27475 \\
C & -0.25840 & 1.94027 & 0.82260 \\
C & 0.27510 & 1.88920 & 2.13149 \\
C & 0.39597 & 3.01746 & 2.94893 \\
C & -0.05833 & 4.25076 & 2.47102 \\
C & -0.60974 & 4.33957 & 1.19056 \\
C & -0.69421 & 3.20013 & 0.38108 \\
N & 0.69592 & 0.58251 & 2.52990 \\
C & 0.64234 & -0.41574 & 1.57665 \\
N & 1.06163 & -1.55175 & 2.20254 \\
C & 1.37240 & -1.29940 & 3.53531 \\
C & 1.13095 & 0.07562 & 3.75794 \\
C & 1.32122 & 0.61656 & 5.03470 \\
C & 1.77221 & -0.23692 & 6.04723 \\
C & 2.02467 & -1.59705 & 5.80996 \\
C & 1.82148 & -2.15128 & 4.54298 \\
C & 1.15115 & -2.86790 & 1.60232 \\
C & 1.57277 & 0.86613 & -1.12299 \\
N & 2.90211 & 0.85694 & -0.83676 \\
C & 3.62273 & 1.51774 & -1.83219 \\
C & 2.67917 & 1.94389 & -2.79719 \\
N & 1.43470 & 1.52825 & -2.31803 \\
C & 3.10883 & 2.62045 & -3.94376 \\
C & 4.47751 & 2.86692 & -4.08565 \\
C & 5.40427 & 2.45280 & -3.11659 \\
C & 4.98683 & 1.76830 & -1.97193 \\
C & 3.47643 & 0.25190 & 0.34849 \\
C & 0.10778 & 1.64036 & -2.83188 \\
C & -0.86007 & 0.99858 & -2.02310
    \end{tabular}
\end{table}
\begin{table}[!t]
    \centering
    \begin{tabular}{ c c c c }
C & -2.18738 & 1.08116 & -2.46920 \\
C & -2.53593 & 1.75933 & -3.64389 \\
C & -1.55142 & 2.38449 & -4.41062 \\
C & -0.21469 & 2.32886 & -4.00287 \\
H & 5.70542 & 1.44575 & -1.21734 \\
H & 6.46567 & 2.66809 & -3.25677 \\
H & 4.83003 & 3.39480 & -4.97434 \\
H & 2.41744 & 2.94786 & -4.71485 \\
H & -2.98249 & 0.60681 & -1.88828 \\
H & -3.58182 & 1.79880 & -3.96125 \\
H & -1.81480 & 2.91604 & -5.32777 \\
H & 0.54353 & 2.82693 & -4.60155 \\
H & 4.56244 & 0.16425 & 0.22218 \\
H & 3.05526 & -0.75016 & 0.48901 \\
H & 2.00197 & -3.20986 & 4.35035 \\
H & 2.37655 & -2.23156 & 6.62596 \\
H & 1.92727 & 0.16996 & 7.04875 \\
H & 1.11601 & 1.65961 & 5.25875 \\
H & -1.11394 & 3.28957 & -0.62375 \\
H & -0.96586 & 5.30355 & 0.81663 \\
H & 0.02850 & 5.13848 & 3.10205 \\
H & 0.85115 & 2.96645 & 3.93492 \\
H & 2.16891 & -3.27153 & 1.72372 \\
H & 0.91581 & -2.78568 & 0.53725 \\
H & -5.77514 & -0.66905 & 1.69364 \\
H & -7.13794 & -2.67275 & 1.05614 \\
H & -6.16764 & -4.44300 & -0.38468 \\
H & -3.85340 & -4.29802 & -1.21524 \\
H & 1.92977 & -3.83701 & -3.24496 \\
H & -0.11652 & -5.25503 & -3.02922 \\
H & -2.05851 & -4.42903 & -1.74828 \\
H & -4.12756 & 1.37436 & 1.10575 \\
H & -2.38106 & 1.41272 & 1.48641 \\
H & 2.02600 & -1.62084 & -2.15665 \\
H & 3.26375 & 0.85636 & 1.24229 \\
H & 0.43810 & -3.55693 & 2.08159 \\
H & -3.52010 & 0.55815 & 2.58191
    \end{tabular}
\end{table}
\FloatBarrier
\subsubsection{Product configuration}
\begin{table}[!h]
    \centering
    \begin{tabular}{ c c c c }
C & -3.21387 & 0.70955 & -1.04295 \\
N & -2.96372 & -0.42896 & -0.18535 \\
C & -1.74006 & -0.78335 & 0.29812 \\
N & -1.95467 & -1.85903 & 1.13096 \\
C & -3.31533 & -2.16538 & 1.20528 \\
C & -3.96132 & -1.23958 & 0.35299 \\
C & -4.06894 & -3.10429 & 1.91637 \\
C & -5.45667 & -3.09955 & 1.73895 \\
C & -6.08716 & -2.18388 & 0.88249 \\
C & -5.34371 & -1.23170 & 0.17871 \\
C & -0.81179 & -2.37556 & 1.81287 \\
C & 0.35525 & -1.59902 & 1.62618 \\
C & 1.51125 & -2.04461 & 2.28181 \\
C & 1.51133 & -3.20017 & 3.07309 \\
C & 0.34491 & -3.95548 & 3.21365 \\
C & -0.83080 & -3.54609 & 2.57600 \\
Ir & 0.14237 & 0.03340 & 0.31124 \\
C & 2.11810 & 0.71875 & 0.58133 \\
C & 3.08923 & 0.07874 & -0.22228 \\
C & 4.42866 & 0.47691 & -0.25852 \\
C & 4.84143 & 1.53084 & 0.56280 \\
C & 3.91665 & 2.16808 & 1.39284 \\
C & 2.57443 & 1.76833 & 1.39071 \\
N & 2.56385 & -0.99186 & -1.00667 \\
C & 1.19659 & -1.15556 & -0.99044 \\
N & 0.94004 & -2.26437 & -1.74020 \\
C & 2.12485 & -2.83426 & -2.20133 \\
C & 3.17842 & -2.01701 & -1.72972 \\
C & 4.50563 & -2.35716 & -2.00825 \\
C & 4.73939 & -3.50663 & -2.77074 \\
C & 3.68680 & -4.30548 & -3.24268 \\
C & 2.35693 & -3.98096 & -2.95794 \\
C & -0.36867 & -2.85249 & -1.92750 \\
C & -0.09478 & 1.79392 & -0.71698 \\
N & 0.27873 & 2.28141 & -1.93301 \\
C & 0.09142 & 3.66100 & -1.99088 \\
C & -0.45255 & 4.04897 & -0.74398 \\
N & -0.56232 & 2.87202 & 0.00066 \\
C & -0.72931 & 5.39593 & -0.49128 \\
C & -0.46412 & 6.32017 & -1.50749 \\
C & 0.06875 & 5.92447 & -2.74379 \\
C & 0.36016 & 4.58159 & -3.00157
    \end{tabular}
\end{table}
\begin{table}[!t]
    \centering
    \begin{tabular}{ c c c c }
C & 0.90915 & 1.50087 & -2.97579 \\
C & -0.94800 & 2.62035 & 1.35207 \\
C & -0.65823 & 1.30623 & 1.78642 \\
C & -0.99342 & 1.00041 & 3.11274 \\
C & -1.58889 & 1.94430 & 3.95905 \\
C & -1.88428 & 3.22499 & 3.48614 \\
C & -1.56801 & 3.57072 & 2.16819 \\
H & 0.79322 & 4.26962 & -3.95290 \\
H & 0.26869 & 6.67472 & -3.51167 \\
H & -0.67368 & 7.37674 & -1.32694 \\
H & -1.12261 & 5.73508 & 0.46331 \\
H & -0.78144 & -0.00122 & 3.49366 \\
H & -1.83183 & 1.67533 & 4.99051 \\
H & -2.36299 & 3.96108 & 4.13632 \\
H & -1.82547 & 4.56084 & 1.79873 \\
H & 0.54197 & 1.82531 & -3.96036 \\
H & 0.65639 & 0.44681 & -2.82149 \\
H & 1.53486 & -4.60906 & -3.30346 \\
H & 3.90907 & -5.19933 & -3.82919 \\
H & 5.76978 & -3.78996 & -2.99557 \\
H & 5.34093 & -1.77045 & -1.63593 \\
H & 1.86186 & 2.28830 & 2.03581 \\
H & 4.24013 & 2.98954 & 2.03841 \\
H & 5.88569 & 1.85173 & 0.54464 \\
H & 5.14856 & 0.00353 & -0.92322 \\
H & -1.12394 & -2.09399 & -1.69889 \\
H & -0.51181 & -3.71089 & -1.25215 \\
H & -5.82960 & -0.50386 & -0.47283 \\
H & -7.17347 & -2.20863 & 0.76990 \\
H & -6.06110 & -3.82481 & 2.28802 \\
H & -3.61016 & -3.81204 & 2.60214 \\
H & 2.42954 & -3.51716 & 3.57502 \\
H & 0.34258 & -4.86838 & 3.81384 \\
H & -1.72892 & -4.15299 & 2.66989 \\
H & -3.99228 & 0.45969 & -1.77811 \\
H & -2.28588 & 0.96075 & -1.56664 \\
H & 2.43595 & -1.47305 & 2.17019 \\
H & 2.00507 & 1.61029 & -2.94039 \\
H & -0.48605 & -3.18586 & -2.96903 \\
H & -3.53674 & 1.58277 & -0.45393
    \end{tabular}
\end{table}
\FloatBarrier
\subsection{[3+2]-Cycloaddition reaction of 9-(2-azidoethyl)-10-(but-3-yn-1-yl)anthracene}
\subsubsection{Reactant configuration}
\begin{table}[!h]
    \centering
    \begin{tabular}{ c c c c }
	C & -5.15792 & -0.78667 & 0.43715 \\
	C & -5.58535 & 0.41936 & -0.05368 \\
	C & -4.68283 & 1.51356 & -0.13724 \\
	C & -3.38146 & 1.36207 & 0.26598 \\
	C & -2.88544 & 0.11903 & 0.78531 \\
	C & -3.81072 & -0.99189 & 0.88586 \\
	H & -6.61800 & 0.53761 & -0.38928 \\
	H & -5.02502 & 2.47369 & -0.53045 \\
	H & -2.70893 & 2.21445 & 0.18151 \\
	H & -5.86096 & -1.61582 & 0.46799 \\
	C & -1.53805 & -0.03325 & 1.18976 \\
	C & -1.10224 & -1.27363 & 1.71342 \\
	C & -2.02964 & -2.38124 & 1.82087 \\
	C & -3.37675 & -2.23019 & 1.41413 \\
	C & 0.24198 & -1.47493 & 2.17379 \\
	C & 0.66147 & -2.67494 & 2.68652 \\
	C & -0.24278 & -3.76703 & 2.77558 \\
	C & -1.53990 & -3.61854 & 2.3582 \\
	H & 0.94619 & -0.64647 & 2.14206 \\
	H & 1.69048 & -2.78982 & 3.03420 \\
	H & 0.09416 & -4.72178 & 3.18567 \\
	H & -2.21510 & -4.46807 & 2.44797 \\
	C & -0.59579 & 1.14756 & 1.12523 \\
	C & -0.69534 & 2.07222 & 2.35323 \\
	H & -0.79866 & 1.75646 & 0.23237 \\
	H & 0.44382 & 0.81463 & 1.02053 \\
	C & -4.32566 & -3.40220 & 1.52896 \\
	C & -4.22165 & -4.41216 & 0.35883 \\
	H & -5.36347 & -3.05749 & 1.60806 \\
	H & -4.13080 & -3.95074 & 2.46119 \\
	C & -4.59694 & -3.86199 & -0.94128 \\
	H & -4.86367 & -5.28249 & 0.58045 \\
	H & -3.18899 & -4.79486 & 0.29831 \\
	C & -4.91430 & -3.38871 & -2.00989 \\
	N & -0.33104 & 1.40568 & 3.60938 \\
	H & -1.71276 & 2.49497 & 2.42485 \\
	H & 0.00842 & 2.90907 & 2.23957 \\
	N & -1.19984 & 0.75690 & 4.18546
    \end{tabular}
\end{table}
\begin{table}[!h]
    \centering
    \begin{tabular}{ c c c c }
H & -5.18981 & -2.96479 & -2.95639 \\
N & -1.91308 & 0.14166 & 4.82329
    \end{tabular}
\end{table}
\FloatBarrier
\subsubsection{Product configuration}
\begin{table}[!h]
    \centering
    \begin{tabular}{ c c c c }
C & -5.43629 & -0.87306 & 0.97441 \\
C & -5.90975 & 0.36890 & 0.63406 \\
C & -5.11516 & 1.52447 & 0.86837 \\
C & -3.87870 & 1.40722 & 1.45151 \\
C & -3.37119 & 0.14170 & 1.87997 \\
C & -4.17104 & -1.03761 & 1.61798 \\
H & -6.89533 & 0.47342 & 0.17558 \\
H & -5.49074 & 2.50773 & 0.57815 \\
H & -3.27550 & 2.29938 & 1.62590 \\
H & -6.04610 & -1.75899 & 0.78808 \\
C & -2.05694 & -0.02688 & 2.37326 \\
C & -1.73941 & -1.23250 & 3.04931 \\
C & -2.52195 & -2.41180 & 2.75963 \\
C & -3.58603 & -2.30701 & 1.82995 \\
C & -0.55868 & -1.38139 & 3.83908 \\
C & -0.18309 & -2.60216 & 4.34924 \\
C & -0.94871 & -3.76237 & 4.05784 \\
C & -2.06993 & -3.66747 & 3.26743 \\
H & 0.03211 & -0.49336 & 4.07253 \\
H & 0.70283 & -2.68160 & 4.98366 \\
H & -0.64813 & -4.72860 & 4.46890 \\
H & -2.65787 & -4.56120 & 3.05058 \\
C & -0.92821 & 0.68660 & 1.67646 \\
C & -0.51730 & -0.23557 & 0.48406 \\
H & -0.03500 & 0.82506 & 2.30364 \\
H & -1.21818 & 1.66553 & 1.27490 \\
C & -3.61976 & -3.30175 & 0.69962 \\
C & -2.38349 & -2.94374 & -0.20858 \\
H & -4.54162 & -3.23153 & 0.10597 \\
H & -3.50854 & -4.34769 & 1.02199 \\
C & -2.39198 & -1.54981 & -0.77853 \\
H & -2.35456 & -3.65206 & -1.05288 \\
H & -1.47058 & -3.11354 & 0.37945 \\
C & -3.23440 & -1.04374 & -1.75549 \\
N & -1.59855 & -0.46479 & -0.47233 \\
H & -0.18303 & -1.20839 & 0.86529 \\
H & 0.31015 & 0.22583 & -0.07449
    \end{tabular}
\end{table}
\begin{table}[!h]
    \centering
    \begin{tabular}{ c c c c }
N & -1.95302 & 0.60391 & -1.21819 \\
H & -4.03310 & -1.55641 & -2.28769 \\
N & -2.93724 & 0.25965 & -1.98581
    \end{tabular}
\end{table}
\FloatBarrier
\subsection{Rotational isomerisation of 9,9'-bianthracene}
\subsubsection{Reactant configuration}
\begin{table}[!h]
    \centering
    \begin{tabular}{ c c c c }
C & -2.68605 & 0.78807 & 1.16242 \\
C & -2.58108 & 2.15674 & 0.70917 \\
C & -1.59264 & 2.49714 & -0.22287 \\
C & -0.69933 & 1.54314 & -0.72566 \\
C & -0.79626 & 0.17340 & -0.27339 \\
C & -1.79094 & -0.18482 & 0.66377 \\
C & -3.70802 & 0.46604 & 2.11302 \\
C & 0.31054 & 1.88567 & -1.68040 \\
C & 1.17700 & 0.93884 & -2.16327 \\
C & 1.08442 & -0.41231 & -1.71624 \\
C & 0.13041 & -0.78215 & -0.80193 \\
C & -1.90326 & -1.60015 & 1.12782 \\
C & -2.70297 & -2.51371 & 0.40531 \\
C & -2.82164 & -3.87946 & 0.86431 \\
C & -2.14101 & -4.27632 & 2.02181 \\
C & -1.34284 & -3.38216 & 2.74596 \\
C & -1.21708 & -2.01582 & 2.29055 \\
C & -3.40904 & -2.13457 & -0.78164 \\
C & -4.17698 & -3.04042 & -1.46871 \\
C & -4.29435 & -4.38666 & -1.01335 \\
C & -3.6359 & -4.79079 & 0.11938 \\
C & -0.39281 & -1.12244 & 3.04810 \\
C & -4.56433 & 1.42831 & 2.58574 \\
C & -4.45807 & 2.77831 & 2.13866 \\
C & -3.49514 & 3.12862 & 1.22714 \\
C & 0.25793 & -1.54676 & 4.17891 \\
C & 0.12865 & -2.89361 & 4.62937 \\
C & -0.64755 & -3.78248 & 3.93098 \\
H & -1.51725 & 3.53216 & -0.56697 \\
H & 0.37418 & 2.92307 & -2.01879 \\
H & 1.94236 & 1.21470 & -2.89288 \\
H & 0.0631 & -1.81642 & -0.46263 \\
H & -2.23451 & -5.30907 & 2.36836 \\
H & -3.32164 & -1.10653 & -1.13524 \\
H & -4.91037 & -5.09401 & -1.57366 \\
    \end{tabular}
\end{table}
\begin{table}[!h]
    \centering
    \begin{tabular}{ c c c c }
H & -3.71983 & -5.82090 & 0.47488 \\
H & -4.70260 & -2.73148 & -2.37518 \\
H & 1.78083 & -1.15724 & -2.10829 \\
H & -5.14817 & 3.53118 & 2.52668 \\
H & -3.40761 & 4.16069 & 0.87815 \\
H & -3.79291 & -0.56486 & 2.45865 \\
H & -5.33604 & 1.16114 & 3.31185 \\
H & -0.75211 & -4.81665 & 4.26933 \\
H & 0.65005 & -3.21226 & 5.53493 \\
H & -0.29049 & -0.09139 & 2.70747 \\
H & 0.88091 & -0.84874 & 4.74324
    \end{tabular}
\end{table}
\FloatBarrier
\subsubsection{Product configuration}
\begin{table}[!h]
    \centering
    \begin{tabular}{ c c c c }
C & -0.79626 & 0.17340 & -0.27339 \\
C & -0.69933 & 1.54314 & -0.72566 \\
C & -1.59264 & 2.49714 & -0.22287 \\
C & -2.58108 & 2.15674 & 0.70917 \\
C & -2.68605 & 0.78807 & 1.16242 \\
C & -1.79094 & -0.18482 & 0.66377 \\
C & 0.13041 & -0.78215 & -0.80193 \\
C & -3.49514 & 3.12862 & 1.22714 \\
C & -4.45807 & 2.77831 & 2.13866 \\
C & -4.56433 & 1.42831 & 2.58574 \\
C & -3.70802 & 0.46604 & 2.11302 \\
C & -1.90326 & -1.60015 & 1.12782 \\
C & -2.70297 & -2.51371 & 0.40531 \\
C & -2.82164 & -3.87946 & 0.86431 \\
C & -2.14101 & -4.27632 & 2.02181 \\
C & -1.34284 & -3.38216 & 2.74596 \\
C & -1.21708 & -2.01582 & 2.29055 \\
C & -3.40904 & -2.13457 & -0.78164 \\
C & -4.17698 & -3.04042 & -1.46871 \\
C & -4.29435 & -4.38666 & -1.01335 \\
C & -3.63590 & -4.79079 & 0.11938 \\
C & -0.39281 & -1.12244 & 3.04810 \\
C & 1.08442 & -0.41231 & -1.71624 \\
C & 1.17700 & 0.93884 & -2.16327 \\
C & 0.31054 & 1.88567 & -1.68040 \\
C & 0.25793 & -1.54676 & 4.17891 \\
C & 0.12865 & -2.89361 & 4.62937 \\
C & -0.64755 & -3.78248 & 3.93098 \\
H & -1.51725 & 3.53216 & -0.56697
    \end{tabular}
\end{table}
\begin{table}[!h]
    \centering
    \begin{tabular}{ c c c c }
H & -3.40761 & 4.16069 & 0.87815 \\
H & -5.14817 & 3.53118 & 2.52668 \\
H & -3.79291 & -0.56486 & 2.45865 \\
H & -2.23451 & -5.30907 & 2.36836 \\
H & -3.32164 & -1.10653 & -1.13524 \\
H & -4.91037 & -5.09401 & -1.57366 \\
H & -3.71983 & -5.82090 & 0.47488 \\
H & -4.7026 & -2.73148 & -2.37518 \\
H & -5.33604 & 1.16114 & 3.31185 \\
H & 1.94236 & 1.21470 & -2.89288 \\
H & 0.37418 & 2.92307 & -2.01879 \\
H & 0.06310 & -1.81642 & -0.46263 \\
H & 1.78083 & -1.15724 & -2.10829 \\
H & -0.75211 & -4.81665 & 4.26933 \\
H & 0.65005 & -3.21226 & 5.53493 \\
H & -0.29049 & -0.09139 & 2.70747 \\
H & 0.88091 & -0.84874 & 4.74324 \\
\\\\\\\\\\\\\\\\\\\\\\\\\\\\\\\\\\\\\\\\\\\\\\\\\\\\
    \end{tabular}
\end{table}

\end{document}